"We Think That We Think Clearly, But That's Only Because We Don't Think Clearly": Brian Josephson on Mathematics, Mind, and the Human World*

About five years ago, Andrew Robinson, who has written quite a bit about Tagore, sent me the Einstein-Tagore discussion (Einstein 1931) and asked me for my comments on it. It seemed to me, as has been said previously, that they were rather talking past each other. Einstein was not understanding what Tagore was saying. The point at issue was whether we have an objective world or a human world. Einstein thought that there was objective reality and that Tagore's position was nonsense. I think my comment at the time was that Einstein was not appreciating how much the processes of construction, engaged in by our senses and minds, affect what we see and what our science consists of. I'll come back to that point later. But anyway I think it's because of that vague comment that I found myself invited here and trying to make sense of this issue. I've been explaining at various times that I don't really understand Tagore and I find what is written on these subjects pretty confusing. But I have been attempting to make sense of it. So what I'm going to do is talk a bit about the relationships between what Tagore seems to be saying and the kind of approach or the kind of problems I've been working on myself.

As I said, the question is whether we have an objective world or a human world, and Tagore's position was that essentially everything is human, everything we know about is human. I'd like to start with something which perhaps not much attention has been given to—the mathematical side. It just so happens that on my way here I looked in the airport bookshop at Heathrow to find something to read, and lo and behold there was a book there called The Mathematical Experience, by Davis and Hersh (1980), evidently put there to assist me with my coming lecture. But let me first, before I say what's in that book, quote from the Tagore-Einstein Dialogue.

Tagore was saying that "truth is the perfect comprehension of the universal mind" and that's the thing I should perhaps just say a bit about. Tagore talks about the individual minds and the universal mind. The universal mind is like a perfected version of the human mind. It's the ideal version of it, I believe. So “individuals approach” the abilities of the universal mind “through our own mistakes and blunders,” our “experience,” and so on. Einstein objected to this in the case of say mathematics. He said, “I cannot prove that scientific truth must be conceived as a truth that is valid independent of humanity, but I believe it firmly. I believe, for instance, that the Pythagorean theorem in geometry,” about the square on the hypotenuse, “states something that is approximately true independent of the existence of man.” So the question was raised, how much mathematical truth is independent of man, and how much is human construction.

Going back to The Mathematical Experience, what a lot of the book discusses is about various happenings in the history of mathematics, what happens when new ideas emerge and people work them through. They also discuss attempts to understand what mathematics is. Because the curious thing is that, once one is trained in mathematics, one does the mathematics fairly automatically. Well, sometimes it's hard work trying to understand something, but one simply feeds the problem into one's mind and then the understanding emerges. But if you ask what is going on, what is mathematics, then this is a very difficult problem. Every approach seems to have some difficulties and to be in some ways not reflecting reality. Davis and Hersch discuss three approaches, or three theories about what mathematics is: Platonism, constructivism, and formalism.

Formalism is the idea that Hilbert introduced in trying to get a proper foundation of mathematics—philosophers like Frege as well. The idea is that mathematics is proving theorems from axioms. So you might think you would simply state what your axioms are, then formalize your processes of deduction. Then you could go through mechanically, or a computer could go through verifying the proof, and that is mathematics. Well the difficulty of this is that, for a start, that we don't really do things that way. There are lots of gaps in the arguments. Then there's the question of where do the axioms come from, because we don't actually play mathematics as a game like chess. The general belief is that we adopt axioms because they are true in some sense. So formalism seems to leave behind discussions of truth and meaningfulness, which seem so important. Furthermore the formal approach runs into trouble, as Gödel showed, since it turns out that axiom systems cannot encompass every true theorem.

The constructivist approach was adopted by people who were doubtful about whether some mathematical arguments were correct. Constructivists say you should not accept the kind of argument which proves that something exists but doesn't tell you how you would find it. People got quite some way using this approach where you only use explicit arguments; you don't allow reductio ad absurdum arguments because there's something a little suspect about them. That doesn't seem to encompass the world of mathematics either and again doesn't agree with many mathematicians' intuitions.

Then the third approach is the Platonic idea or Platonism. Gödel was one of those who advocated this approach. Gödel, like many mathematicians, thought that we have direct contact with mathematical truth: there's some domain of knowledge we could contact. To the extent we could expand our minds to experience that domain, we would know what is true and what is not. At least this would tell us where the axioms come from. Thus we could determine truth or falsity even in some cases where we can't seemingly prove whether a result is true or not—the continuum hypothesis is one of these. One of the things it points out in this book is that although we can't by axioms prove or disprove the continuum hypothesis, we might someday get enough insight into what this particular theorem means that we would recognize whether it was true or not. So, the Platonic concept is that really there is a mathematical truth independent of man and all we do is observe it, just as we observe a physical phenomenon.

All of these approaches have difficulties. Let me discuss what kinds of difficulties there are with Platonism. Partly we seem actually to be fallible. There seems to be something wrong with the idea that we can contact truth—for what we take to be truth, changes. So Davis and Hersh advocate an idea which is actually much closer to Tagore's and it stems from the philosophy of Lakatos, who is in the tradition of Popper. Popper took a simple idea that we arrive at the notion of truth from the fact that ideas can be falsified; so the way science evolves is that people make conjectures, find ways of testing them, and some of them are falsified—but those that remain become the body of science. Well, Lakatos took this idea in a slightly different direction in that he said when your hypothesis is disconfirmed, you just don't throw it away. You can rescue it by changing your assumptions. So you've got a complicated situation where science occasionally has to sidestep slightly and go in a different direction. Still, you approach some kind of truth.

Anyway, Davis and Hersh treat mathematics just like any other ordinary human experience. Their argument as to why mathematics seems to be objective is that people talk to each other; they explain what their ideas mean. This process of communication leads everyone to get mental states with the same sort of structure, and because their mental states have the same sort of structure, they will come to the same conclusions as to what is true or not. Davis and Hersh go through all sorts of arguments to make the point that really mathematics is like this. We don't automatically understand something. We talk it over and gradually things become clearer as our minds get at it. We are trained to think in the right way about a new mathematical idea. And only people with brains that are sufficiently trainable will be able to understand the ideas in the first place. So they have this sort of fallibilist approach and it's an approach to mathematics which reconciles the apparent universality and objectivity, but also sees it as a human activity. So what they arrive at—and I think it's a very compelling argument, detailed with lots of examples of how mathematicians actually work—is a Tagorian viewpoint that mathematics is a particular kind of human activity which comes out with something apparently universal. But there is nothing especially privileged about mathematics.

This recalls an argument I had at a following a conference in Finland a few years ago with a philosopher, Rachel Waugh, who gave a talk (Waugh 1995) insisting that meaning came before truth. I said isn't mathematics different, isn't that real truth? I found I was unable to produce a really convincing argument for mathematical truth being real. In the end I produced my own undestanding of the situation on the basis of an assertion that mathematics is just like knitting. What's it got to do with knitting? Well knitting is a set of practices designed to produce something like a sweater which you can wear and which protects you against the cold. A valid knitting 'theorem' would detail a process which produces something that works, like a sweater resistant to the wind. In the same way, a valid mathematical idea such as that of number or mathematical induction is something that works and is resistant to the winds of skepticism. This strictly has to be done in cases like the treatment of infinity. There one has to work quite a way to find the garments or the sort of ideas which are resistant to the winds of skepticism. Anyway, from this point of view the only real difference between mathematics and knitting is that mathematical truths are much wider in their scope than the truths of knitting. So there is something special about mathematics, but nevertheless it's a particular subset of all possible activities.

Okay, so that's the human sense of mathematical truth. Let me now move on from there to physical truth. I think here the case for science revealing human truths is much easier to make because science changes from era to era, particularly say at the foundation of high energy physics. Clearly there is some human activity which is trying to make sense of nature, trying to find the truth of nature. But it is always provisional and over the course of time you may see things in a completely different way, as science advances. So however much people think they're coming to a theory of everything—and these things always seem quite doubtful anyway as evidence comes up against them—these just seem to be provisional truths which are derivative of a particular way in which one approaches the study of the universe at some particular time. Indeed, beyond what we do as science there may be an indefinitely large part of reality that we know nothing about. Moreover, what we do perceive is very much structured by our tools. For example, the fact that I perceive a visual scene as people, chairs and so on, is certainly very derivative upon the fact that I have sense organs which are especially adapted to see these kinds of things. Also, the particular scientific theories we get are very much a result of particular mathematical and experimental tools we used. So we're really being very something-centric; it's a very parochial view we have. We imagine that this is how things really are. In fact, it’s just a view of nature from a particular standpoint of what we know at this particular era.

It's very interesting to ask why people get the idea that science is giving a complete view of nature. I think it's related to what Jonathan Shear was saying yesterday, or an extension of it, about the way children of a particular age get the concept of an object. When you cover something like a ball with a cloth, before a certain age children act as if it no longer exists, but later on they act as if it's still there and they try to remove the cloth to get at the object. Now clearly it is adaptive to have this object concept. This concept says there's a reality beyond what you know. It is adaptive because it often works. The thing can be recovered and hasn't gone away forever. As the philosopher Merleau-Ponty pointed out (Dillon 1988), this is what seems to be behind science. We have an objectifying instinct which causes us to view the universe in objective terms and do what we can with it. This is a very adaptive kind of thing because to the extent that this objectivization of reality is the case we can form plans for dealing with the universe. For example, if a timetable tells us that a bus will be present at a particular time at a particular place, we can go there and catch the bus rather than going by chance. That's one form of objectivization. So it is an important human capacity to represent the universe as if it were what we thought it was. It takes a lot of load out of our theorizing if we have this capacity to think of the universe as being so and so. But, of course, the universe never actually is the same as what we think it is. What we think it is is only an approximation, even if we define what the correspondence is.

So, in other words, it does seem Tagore is right that we have this human vision which defines what the world is for us. We have an inability, or at least a difficulty, in looking beyond it and seeing that this might not be the whole truth.

I thought of an aphorism once to summarize this which I have on my blackboard. It says that "We think that we think clearly, but that's only because we don't think clearly." You see the natural automatic thing is to think that all these thoughts are really representing how things are. But the world was not that simple. We're not seeing things clearly enough to recognize that things are not that clear.

I want to go into some different kinds of issues now. Let me introduce this by talking about something that is very familiar to us in physics and that's the property of sound waves. Because here this is a case where we have a spuriously simple mathematics. This is in fact an illustration of what I was talking about, where the science of one era is superceded by the science of a later era. Because a sound wave is something which is just like a wave form that propagates for a certain velocity in accordance with the wave equation. But we know that that's not really what's going on. There's another level, but let me not go to the ultimate level. Let me go to classical physics according to which the gas (in which the sound wave moves) is made up out of a collection of particles which bounce off each other. We know from this kinetic theory of gases that there's a deeper underlying reality, a more complicated reality which is underneath the simpler reality. The message we can learn here is that there may be a more complicated reality underneath the reality that we study by physics.

There's an interesting point I can make here as a halfway stage to where I want to get to in a moment. Suppose we just dump molecules randomly into a container. That would then correspond to a very turbulent kind of motion which might not support sound waves. What causes sound waves to be possible is that, first of all, the gas organizes itself so it is in some kind of rough equilibrium. It is that equilibrium state which supports the sound waves. So the manifest, orderly phenomenon of sound waves comes out of an unmanifest kind of order, the background equilibrium system. It's been recognized in the last few years that that kind of thing happens and is a sort of a correction to how one normally thinks about science. This is the phenomenon known as emergence. It is something which applies to all sorts of complex systems like the weather, biology, and so on. It's now being recognized that if we have a complicated system, it will go along obeying one kind of law and then all of a sudden there will be a switch and a different kind of phenomenon will emerge out of it. And it will do so in an essentially unpredictable way. So there's a certain random element to what we observe. There's a more fundamental background which organizes itself and after it has organized itself some various phenomena will emerge. So in all these cases the behavior you observe is not the fundamental behavior. It's contingent on what's going on in the deeper level.

So I think the question we need to ask now is—can we say anything about this deeper level. All these arguments I've talked about here suggest that what we observe through science, this human activity, is not the most fundamental activity, not the most fundamental phenomenon possible. Here we get to things which I don't know too much about so I can only speculate and perhaps others will speculate with me in the discussion. Tagore talks about there being an "absolute truth," Brahman, and this may be connected with the background order out of which manifest phenomena emerge. One of the points Tagore makes, and all the mystical traditions, is that you cannot discuss this background order in the same sort of way. You can know it and experience it, but you cannot analyze it in a scientific manner, as this thing is beyond our usual rational activities through which we try to understand. One thing I have discussed as part of the mind/matter unification project we have at Cambridge is the phenomenon of music. I don't know perhaps where to place that. But the suggestion that came out of this research, which I've been doing in collaboration with a musicologist, Tethys Carpenter (Josephson and Carpenter 1996), is that there are certain fundamental forms which have an activising effect on life. We can study what these forms are in music, and perhaps you can again study the forms by meditation. Then there would be some different kind of science to the kind we have now, based on subjective experience, though, whether we are going to call that science or not is unclear. Are we getting beyond Tagore's "human truth" by these means? I don't know what Tagore would have said about that.

So, to summarize, I think one can certainly justify by a number of arguments Tagore's point that our science as a human activity and our universe also may be a human universe, possibly something which came into existence at the time of the big bang. Tagore has this picture that you have a strongly interconnective system, like a solid but that's composed of people and a material universe. There may be some special system which science has become most aware of. But there may be something else beyond that which will be the task of future science to try and understand.[1]

*Talk delivered 20 September 1998, Storrs, CT. Transcription by Anne Theriault. Edited by Patrick Colm Hogan.

[1] Possible clarification of the situation is given by a more detailed reading of Waugh's paper, where she writes of "representational contents which contain the germs of understanding but which do not make truth claims and whose meanings are not given by truth conditions". The forms in music in the paper by Carpenter and myself (where we wrote of composer's "intuitive ability to be aware of the creative potentials of particular sounds even when considered in their most elementary forms) are in some respects similar. Here are things differing from the ordinary

concept of truth; perhaps they can be characterized as “truthful experiences”, emerging from Brahman: truthful in the light of their positive potential.